A simple geometric description of T-duality is given by identifying the
 cotangent bundles of the original and the dual manifold. Strings
 propagate naturally in the cotangent bundle and the original and
 the dual string phase spaces are obtained by different projections.
 Buscher,s transformation follows readily and it is literally projective.
 As an application of the formalism, we prove that the duality is 
 a symplectomorphism of the string phase spaces.

\documentstyle[12pt]{article}
\begin{document}
\def\om{\omega}
\def\omt{\tilde{\omega}}
\def\ti{\tilde}
\def\o{\Omega}
\def\t{T^*M}
\def\vt{\tilde{v}}
\def\ot{\tilde{\Omega}}
\def\otwo{\omt \wedge \om}
\def\owot{\om \wedge \omt}
\def\w{\wedge}
\def\mt{\tilde{M}}
\def\ss{\subset}
\def\pit{\tilde{\pi}}
\def\tpm{T_{P} ^* M}
\def\al{\alpha}
\def\alt{\tilde{\alpha}}
\def\inop{{\int}^{P}_{P_{0}}{\om}}
\def\th{\theta}
\def\tht{\tilde{\theta}}
\def\inox{{\int}^{X}{\om}}
\def\inotx{{\int}^{X}{\omt}}
\def\st{\tilde{S}}
\def\l{\lambda}
\def\p{{\bf{p}}}
\def\pb{{\p}_{b}(t,u)}
\def\pbm{{\p}_{b}}
\def\del{\partial}
\def\l2{\Lambda^2}
\def\be{\begin{equation}}
\def\ee{\end{equation}}
\def\ej{{\bf E}}
\def\ed{{\bf E}^\perp}
\def\d{\cdot}

\begin{titlepage}
\begin{flushright}
CERN-TH.7490/94
\end{flushright}
\vskip 1cm
\begin{center}
{\large\bf Strings in Spacetime Cotangent Bundle \\  and T-duality} \\
\vskip 1cm 
{\bf C. Klim\v c\'\i k
\footnote{On leave from Charles University, Prague}} \\
\vskip 0.3cm
  {\it Theory Division CERN, CH-1211 Geneva 23,
Switzerland} \\
\vskip 0.5cm {\small and} \\
\vskip 0.5cm
 {\bf P. \v Severa}\\
\vskip 0.3cm
 {\it Dept. of Theoretical Physics, Charles
University,} \\ {\it V Hole\v sovi\v ck\'ach 2, CZ-18000 Prague,
 Czech Republic} \\ 
\end{center}
\vskip 1cm  
\begin{abstract}
  
A simple geometric description of T-duality is given by identifying
 the cotangent bundles of the original and the dual manifold. Strings
 propagate naturally in the cotangent bundle and the original and the
 dual string phase spaces are obtained by different projections.
 Buscher's transformation follows readily and it is literally
 projective. As an application of the formalism, we prove that the
 duality is a symplectomorphism of the string phase spaces.
 
\end{abstract}
\vskip 1cm
CERN-TH.7490/94\\
October 1994
\end{titlepage}

\section{\bf Introduction}

Target space duality keeps attracting the attention of string theorists
(see e.g. \cite{Busch,Duff,RoVe,Kir})
 mainly because it deepens our understanding of the geometry of spacetime
 from the string point of view  and it  is an important tool for
 disentangling the full symmetry structure of the string theory.

Duality is usually derived in the $\sigma$-model context in appropriate
 coordinates, respecting the isometry of the action. The symmetry is
 gauged and the gauge fields are constrained to be trivial by
 introducing a Lagrange multiplier. The latter is understood as
 a new (dual) coordinate and the gauge fields are integrated out
 to end up with the dual $\sigma$-model. The dual metric $\ti G$
 and the skew-symmetric tensor $\ti B$ are then given by Buscher's
 formula.

This approach is not very transparent from the geometrical point of view,
 however. An important conceptual simplification was undertaken in
 \cite{Alv1} where the duality was described in a global geometric
 setting and, following the
suggestion in \cite{GiRa},
 it was interpreted as just a canonical transformation of the phase
 space of the theory \cite{Alv2}.
 Attempting to further clarify the concept, we give a very simple
 global geometric description of duality for a large class of systems 
including the $\sigma$-models. We show that the dual target, its
 cotangent bundle, and the elements of the original and the dual phase
 spaces can be naturally embedded into the cotangent bundle of the
 original target. The relation between the original and the dual
 quantities is now simply given by different projections of invariant
 objects living in the cotangent bundle. Buscher's formula follows
 extremely straightforwardly and it is literally projective.
 As an application of the developed formalism we present a simple proof
 that duality acts as a symplectomorphism  of the original and
 the dual phase spaces, understood as appropriate submanifolds of
 the cotangent bundles of the loop spaces of the targets. The whole
 picture seems quite natural. We may even guess that the cotangent
 bundle will play an important role in developing a natural framework
 for the description of string symmetries. 

The plan of the paper is as follows. In section 2 we identify the
 cotangent bundles of the original and the dual manifolds and compare
 their canonical symplectic structures. In the next part we lift the
 dynamical characteristics of strings into the cotangent bundle and
 by different projections we obtain Buscher's formula. In section 4
 we prove that the duality is a symplectomorphism and in section 5
 we consider $\sigma$-models only, recovering the standard results 
in a very compact way.

\section{\bf Dual symplectic structures on \break the cotangent bundle}

 Consider a manifold $M$ with a global vector field $v$ and a global
 closed 1-form
${\om}$ such that $\om (v)=1$.\footnote{The form $\om$ is locally the
 differential of a coordinate along $v$.}
  Let $\t$ be the cotangent bundle of $M$. The fields $v$ and $\om$ are
 naturally extended to the whole ${\t}$. On ${\t}$ there is the
 canonical symplectic form ${\o}$. We define the dual fields ${\vt}$
 and ${\omt}$ on ${\t}$ in this way:
 \be \omt=\o (.,v)\qquad {\rm and} \qquad \om=\o (\vt ,.)  \ee
 (so that $d\omt=0$ and $\omt(\vt)=1$).

There exists another symplectic form ${\ot}$ on ${\t}$ such that
\be{{\ot}+{\otwo}={\o}+{\owot}}.  \ee
 Then obviously
 \be \om=\ot(.,\vt)\qquad {\rm and} \qquad \omt=\ot(v,.)  \ee
 These relations are manifestly dual to (1). This suggests that ${\t}$
 can be interpreted as the cotangent bundle of some dual manifold ${\mt}$
 whose canonical symplectic form is ${\ot}$. We obtain ${\mt}$ by
 the action of the vector field ${\vt}$ on a hypersurface ${M_{0}{\ss}M}$
 such that ${{\om}|_{M_{0}}=0}$. Then clearly $M_{0} =M \! \cap \! {\mt}$
 and ${{\omt}|_{M_{0}}=0}$. The dual projection ${\pit}$ maps
 $(P\! \in \! M, \al\! \in\! \tpm)\! \in \t$ into $(P_{0} \! \in
 \! M_{0},{\al}(v){\om})\! \in \! \mt$, where ${P_{0}}$ lies on the
 same integral curve of $v$ as $P$ does. Every point
 ${(P,{\al})\! \in \t}$ can be understood as a 1-form on ${\mt}$
 at the point ${{\pit}(P,{\al})=(P_{0},{\al}(v){\om})}$. It associates
, to any vector $t_0$ annihilated by $\omt$, the number $\al (t_0 )$,
 and to $\vt$ the number $\inop$. If the integral curves of $v$ are
 closed then the function ${f(P)={\inop}}$ is multivalued. In this
 case we have to identify ${\t}$ with ${T^*{\mt}}$ factorized by an
 appropriate discrete group.

 There remains to show that Eq. (2) holds. Because ${{\ot}=d{\tht}}$
 (and ${{\o}=d{\th}}$) where ${\th}$ and ${\tht}$ are canonically defined,
 it is sufficient to compare ${\tht}$ and ${\th}$. Every vector
 ${t \! \in T_{X}({\t})}$ can be uniquely written as
 ${t=t_{0}+{\om}(t)v+{\omt}(t){\vt}}$ where
 ${{\om}(t_{0})={\omt}(t_{0})=0}$. Then by definition
 ${{\th}(t_{0})={\tht}(t_{0})}$, ${{\th}({\vt})=0}$, ${{\tht}(v)=0}$,
 ${{\th}(v)={\inotx}}$ and ${{\tht}({\vt})={\inox}}$. Hence
 $${{\tht}(t)={\th}(t)-{\om}(t){\inotx}+{\omt}(t){\inox}}$$ and Eq. (2)
 follows. As the formulae suggest, starting with the dual fibration
 and repeating the procedure will bring us back to the original one.

We shall argue that the framework described above is very well suited 
for the description of the (Abelian) target space duality in string theory.

\section{\bf Strings embedded into $\t$ and Buscher's formula}

Suppose $M$ is a spacetime in which string propagation is governed by
 an action $S$, invariant with respect to the vector field $v$. There
 is no need to assume that $S$ is a ${\sigma}$-model action; we only
 suppose that $S$ is local, reparametrization-invariant
 and depending on the first derivatives of the embedding of the string
 worldsheet into $M$.

 Because $v$ is a symmetry, by Noether theorem there is a closed 1-form
 ${\al}$ (the density of the $v$-component of the momentum of the string)
 on any surface extremizing $S$. Due to this fact, any on-shell string
 can be naturally lifted to $\t$ along $\vt$ so that the form $\omt$
 restricted to the lifted surface gives precisely $\al$.\footnote{
This lift is defined uniquely up to a uniform shift by ${\vt}$.
 We shall discuss the ambiguity in the next section.}

Then we project the lifted surface to ${\mt}$. We shall show that
 there exists a ${\vt}$-invariant action ${\st}$ whose extremal
 surfaces are just the projections to $\mt$ of the lifted surfaces.
 The action ${\st}$ will obey the duality property, namely the lift
 of the dual extremal surface along $v$, such that the dual
 Noether form ${\alt}$ is restricted ${\om}$, coincides with the
 original lift.

All that picture should be refined, however, in the case of closed
 strings. Then the form ${\al}$ need not be exact and the lift of
 the non-contractible loop on the worldsheet may give an open curve
 in ${\t}$. Therefore, we have to identify the points of ${\t}$
 along ${\vt}$ in such a way that
$${n{\oint}_{orbit~ of~ {\vt}}{\omt}={\oint}_{loop}{\al}=p},$$
where $p$ is the $v$-component of the total momentum of the string
 and $n$ is
an integer. As a consequence, the momentum of the string is an integer
 multiple of some minimal momentum and the dual string winds $n$
 times around the orbit of ${\vt}$. The picture holds in the dual
 version, of course.

In order to find the dual action, we have to identify a geometric
 object on $M$ which encodes the original action, can be naturally
 lifted to ${\t}$, and then projected to the dual manifold ${\mt}$.
 Because the action is reparametrization-invariant and depends at
 most on the first derivatives, the Lagrangian $L$ is a function
 of decomposable bivectors $b$ at any point $P$ of $M$ such that
\be{L(\lambda b)=\lambda L(b), \qquad \lambda >0}.   \ee
This object is not convenient for lifting to ${\t}$. Instead, we define
 a bilinear mapping $\p$
\be\pb\equiv {d \over {d \epsilon}}L(b+\epsilon t \w u) 
\vert_{\epsilon=0},\ee
where $t,u\in {T_{P}M}$ and $t \! \w \! b=0$ (the last condition
 means that the vector $t$ lies in the plane of $b$ and in fact
 ensures that the argument of $L$ in (5) is decomposable). Note that
 $\p_{\lambda b}(t,u)={\pb}$ or in other words the mapping ${\p}$
 depends only on the plane in which $b$ lies. Physically speaking,
 ${\pb}$ is the density of the $u$-component of the momentum.\footnote{
We may say that ${\pb}$ defines for any embedding of string into $M$
 a 1-form on the worldsheet with values in 1-form on the target.
 In arbitrary coordinates ${{\zeta}^{\al}}$ on the worldsheet and
 ${X^{\mu}}$ on the target it can be written as ${{\p}_{{\al}{\mu}}=
\del({\cal L} \epsilon_{\al\beta})/{\del}({\del}_{\beta}X^{\mu})}$,
 where ${S={\int}{\cal L}d{\zeta}^{0}d{\zeta}^{1}}$.}

We see that the mapping ${\pb}$ contains the essence of dynamical
 properties of string. Obviously, $\pbm (b)=L(b)$ and
 $\pbm (t,v)={\al}(t)$ where ${\al}$ is the mentioned Noether 1-form.

A bivector ${b=v_{1} \! \w \! v_{2}}$ at a point ${P \! \in \! M}$
 can be naturally lifted  to a decomposable bivector
 ${b^{*}\!\in\!\l2 T_{P}({\t})\!\simeq\!\l2 (T_{P}M+{\tpm})}$:
\be {b^{*}=(v_{1}+\pbm (v_{1},.)) \!\w\! (v_{2}+\pbm (v_{2},.))}.  \ee
We observe the simple formula
\be {2L(b)={\o}(b^{*})}.    \ee
The dual Lagrangian $\ti L$ is defined in such a way that lifting
 $\ti b =\pit (b^* )$ by $\ti L$ gives $b^*$. Then obviously\footnote{
Speaking more exactly, the value of ${\o}$ is the same for any bivector
 on ${\t}$ obtained by acting by the vector fields ${v,{\vt}}$ on
 ${b^{*}}$, because the symplectic form ${\o}$ is ${v,{\vt}}$-invariant.
 This means that we can transport $b^*$ into a point of the dual
 manifold $\mt$ embedded in $\t$ and write the formula (8) there.}
\be 2 \ti L ( \ti b )=\ot (b^*) .              \ee 
From (2), (7) and (8) Buscher's duality transformation follows:
\be{{\ti{L}}({\ti{b}})=L(b)-({\otwo})(b^{*})=L(b)-(\al \w \om)(b)}. \ee
By construction, the dual Lagrangian is ${\vt}$-invariant. In a way,
 it may be interpreted as the Routh function, because $\om$ is the
 differential of the coordinate along the symmetry field and $\al$
 is the corresponding momentum. Later we shall write the formula
 in the familiar $\sigma$-model context. The formula
can be derived also relaxing the condition of reparametrization
 invariance.
However, its derivation is slightly less straightforward.

 We should demonstrate that we obtain, by minimizing the dual action,
 the surfaces in ${\mt}$ projected from the lifted extremal surfaces
 of the original action. First notice that for lifting a surface
 ${F\!\ss\! M}$ we only need ${d{\al}=0}$. Upon gauging the symmetry
 $v$ the variation of the action $ S(F+\epsilon v)-S(F)$ is equal
 to $\int_F d\epsilon\w\al$. It means that ${d{\al}=0}$ iff $F$
 is extremal with respect to arbitrary (non-uniform) variations in
 the direction of $v$. By construction, the same thing is valid for
 dual objects, i.e. the liftable surfaces are exactly those extremizing
 $S$ and ${\ti{S}}$ with respect to the $v$- and ${\vt}$-direction
 respectively. Therefore we shall restrict our attention  to these
 surfaces only. Now from (9) one easily observes
\be {{\ti{S}}({\ti{F}})=S(F)-{\int}_{F^{*}}{\otwo}},             \ee
where ${F^{*}}$ is the common lift  of ${F{\ss}M}$ and
 ${{\ti{F}}{\ss}{\mt}}$.  The difference between the actions
 is an integral of a closed form so it does not feel any variation.

\section{
\bf  Duality as a symplectomorphism of \break string phase space}

By the phase space of a closed string theory we understand the space of
 all classical solutions having the topology of a cylinder. On this
 space there is a natural symplectic form $\o _{Ph}$ coming from the
 action. However, we had to quantize the momentum for the duality
 to make sense. If we fix the momentum then we obtain a hypersurface
 in the phase space. This is not a symplectic space because $\o _{Ph}$
 is not invertible on it. To obtain a symplectic structure one has to
 factorize this hypersurface: one identifies each string $F$ with all
 the strings obtained by tranlating $F$ by the vector field $v$.
 This is (the simplest case of) the Marsden-Weinstein reduction
 \cite{Arn}: if there is a function $p$ on a symplectic space generating
 a vector field $w$ then one sets $p={\rm const.}$ to obtain
 a hypersurface and then factorizes by $w$; the result is a symplectic
 space. This factorisation is perfectly suited for the duality because
 the dual string $\ti{F}$ is only defined up to a shift by $\vt$ and
 does not depend on shifting $F$ by $v$. So, duality is a one-to-one
 mapping of the reductions. There is a pretty physical reason for
 the factorization: if a string has an exact value of the momentum
 then its state does not change if we shift it by $v$.

Now we prove the following proposition: the duality is a
 symplectomorphism of the (reduced) phase spaces.

Proof: Let $LM$ be the loop space of the target $M$. As usual,
 we obtain the phase space from the cotangent bundle $T^*LM$, on
 which there is the canonical symplectic form $\o _{Ph} =d \th _{Ph}$.
 Namely, we identify some submanifold in $T^*LM$ and then factorize
 it appropriately.\footnote{We proceed conceptually as in the case
 of a relativistic particle in a background; in the $\sigma$-model
 case the submanifold is defined by the Virasoro constraints.}
 The construction goes as follows: if we have a string worldsheet $F$
 and a loop $l$ on it, then we define a corresponding element
 $l_F \in T^*_l LM$. To describe how $l_F$ acts on a vector
 $u \in T_l LM$, first realize that $u$ can be thought of as a
 family of vectors $u(X) \in T_X M$ where $X$ runs along $l$. Then
\be l_F (u) \equiv \oint _l \pbm (.,u(X)). \ee
If we take all $l_F$'s for all possible $F$'s we obtain the mentioned
 submanifold of $T^*LM$. Now we identify all $l_F$'s coming from the
 same extremal $F$ and obtain the phase space.

In this framework we can easily prove the proposition. Let $H$ be a 
surface in the original phase space, i.e. a 2-parametric family of 
on-shell strings, and let on each $F \in H$ be a loop $l(F)$.
 Then by (11)
$$\int _H \o _{Ph} = \oint _{\del H} \th _{Ph}
 = \oint\limits_{\bigcup_{F \in \del H} l(F)} \p_{b(F)}(.,.) $$
The last integral is over a closed surface in M. We will prove that
 if $\ti H$ is a corresponding family in $\mt$ (defined up to an
 independent shift of each $\ti F$ by $\vt$) then     
$$ \oint\limits_{\bigcup_{F \in \del H} l(F)}\p =
 \oint\limits_{\bigcup_{\ti F \in \del \ti H} l(\ti F )}\ti{\p} . $$
We will compare the two expressions using the common lifted family
 $H^*$. One immediately checks that if $t_*$ and $u_*$ are vectors
 at a point of a lifted surface $F^*$, $t_*$ tangent to $F^*$ and
 $u_*$ arbitrary, then 
$$\ti{\p}(\ti{t},\ti{u})-\p (t,u)=(\owot)(t_* \w u_*)$$
so that
$$ \oint\limits_{\bigcup_{\ti F \in \del \ti H} l(\ti F )} \ti{\p}-
 \oint\limits_{\bigcup_{F \in \del H} l(F)}\p=
\oint\limits_{\bigcup_{F^* \in \del H^* }l(F^*)}\owot =0$$
because $\owot$ is closed and the closed surface over which we
 integrate is a boundary.

\section{\bf $\sigma$-model and projective transformations}

 In what follows we shall study the duality in the familiar context of
 the non-linear $\sigma$-model. In this case there is a metric $G$ and
 a 2-form $B$ on the manifold. The action $S$ is minus the area of the
 surface plus the integral of $B$ over the surface. We assume the
 surface to be time-like everywhere, i.e. there are two real light-like
 tangent vectors $k,l$ at any point of the surface; we always choose
 both of them lying on the same light cone (future or past). Then
 the area of $k \w l$ is simply $-k\d l$, i.e. 
\be L(k\w l)=G(k,l)+B(k,l)\equiv E(k,l)  . \ee
It means that
\be \p_{k\w l} (k,.)=E(k,.) \qquad {\rm and} \qquad \p_{k\w l} (l,.)
=-E(.,l) . \ee 

 The duality transformation follows directly from Buscher's formula
 (9), but there is a simpler geometric way of deriving it in the
 $\sigma$-model context. Using Eq. (13) and the decomposition
 $T_P (\t)\simeq T_P M+\tpm\simeq T_P\ti M +T_P ^* \ti M$ we
 interpret $k^*=k+\p_{k\w l}(k,.)=k+E(k,.)$ as $\ti k + \ti E (\ti k,.)$,
 and accordingly for $l^*$, thus obtaining the dual bilinear form
 $\ti E =\ti G +\ti B$. That is, we interpret the graph of $E$,
 $\ej=\{t+E(t,.)|t\in T_P M\}$, from the dual point of view as the
 graph of $\ti E$. 

The projective formula for $\ti E$ is self-evident now. One simply
 exchanges $T_P M$ and $T_P \ti M$ by the linear transformation
 $R$ of $T_P(\t)$, $v\leftrightarrow\vt$ and $w\mapsto w$,
 if $\om(w)=\omt(w)=0$. Then $\ti t +\ti E(\ti t,.)
=R(t+E(t,.))$.

Our formalism can be easily extended to the case of $d$ commuting
 symmetries. Then $R$ is an element of a group $O(d,d;{\bf Z})$
 preserving the natural metric on $T_P (\t)$, $(t+\beta)^2=\beta(t)$.

\end{document}